# Studying of refractive index measurements in reflected light

E.A.Tikhonov, V.A.Ivashkin
Institute of physics Nat. Academy Sci., Ukraine, Kiev, 03680, prospect of Science 46,
**etikh@iop.kiev.ua**

Two methods of refractometry in reflected light from optical surface of samples are considered and studied experimentally. Methods are grounded on results of Fresnel theory of concerning light reflectivity at near normal incidence and Brewster angle. Sources of errors for both methods were considered and possibility of measuring of the refractive index with application of laser radiation with accuracy to within 4th sign was shown. Advantages of described methods concerning requirements to preparation of samples to refractive index measurement of solid, thin-film and absorbing materials are scored.

## 1. THE INTRODUCTION

The importance of precise measurements of material index refraction (further IR) for scientific, medical, military instrument making and materials technology is well-known. These measurings play not passing role a chemical analysis and synthesis of new materials. Recently the great attention is given to examination of metamaterials with negative IR [1,2]. The so-called profile refractometry develops promply for measuring of local values of IR fluids in time and the space that is related with application of Sore effect – mass diffusion at a thermal gradient. The coherent optical befocal refractometry is engaged in measuring of IR scattering mediums [3].

Classical refractometrucal procedures and devices are rather full described in D.B.Ioffe's monography [4]. Under widespread approaches which are embodied in refractometrical devices, it is possible to formulate the short inference as embodying of TIR measurings and refraction angle measurings (goniometric methods). The preference to angular methods of IR measuring is donated because the systematic error source of measuring go to mimimum. Development of the termed procedures with application of tunable lasers allows to raise further accuracy of measurings of a IR dispersion not only in the characteristic points of a spectrum, but in a wide spectral range also.

It is possible to guess a series of the reasons why among angular methods of a refractometry methods on the basis of reflected light measuring, in particular, IR measurings on a Brewster angle have not found usefull. The termed method demanding measuring only one angular parametre is characterised by a unique systematic error also. Moreover given method using reflexion of light from the single surface does not depend on a physical state of a material, its shape and, partially, absorption constants and scattering. The numbered advantages of this approach potentially expand measuring possibilities in comparison with the developed methods of TIR and goniometric one. Insuperable restriction and in this method there is a state reflected material surface: this surface cannot be scattering.

The thesis about the aprioristic advantages of a method of measuring of IR on a Brewster angle can raise the doubts at the account of observable infringement of Fresnel theory in a Brewster angle neighbourhood: presence of reflected light in a neighbourhood of an exact Brewster angle even at 100 % of polarisation and zero value of a lateral angle (a corner of a mismatch of planes of incidence and polarisation) [5]. These questions will be discussed in the section featuring embodying of this method more low.

Other possible procedure of measuring of IR on reflected light at normal incident also leans against deductions of Fresnel theory. But in this case there are no doubts in the theory and there are no requirements to accuracy of the angular measuring. However there are rigid requirements to accuracy of power measurings of incident/reflected light (up to 4-5 signs) in order to have the equivalent measureming accuracy of IR. Presence of noise and instability of power for noncoherent sources and multimode lasers of light create notable restriction on accuracy of similar measurings though application of numeral procedures of measuring allows to overcome the termed difficulties partially.

## 2. MEASURING IR AT SMALL ANGLES OF INDICENCE

The given procedure is followed on Fresnel theory for dependence of light reflection from refractive coefficient «n» for 2-x orientations of the linear polarized light concerning a plane of incidence /5/:

$$R(\perp) = \sin^2(\varphi-\psi)/\sin^2(\varphi+\psi) \qquad (1a)$$

$$R(\parallel) = tg^2(\varphi-\psi)/tg^2(\varphi+\psi) \qquad (1b)$$

Here $\varphi$ and $\psi$ are angles of incidence and refraction on a surface of a material with IR «n». At normal slope $\varphi=\psi=0$ quantities of reflection for radiation both polarization are getting equal and spotted only quantity «n». Thus IR it is determined with accuracy, restricted errors of measuring of quantity R:

$$n = (1+\sqrt{R}) / (1-\sqrt{R}) \qquad (3)$$

For some reasons, following from method of incidence and reflected power measuring of laser radiations to realise measuring at strictly normal slope it is almost impossible; at slope under arbitrary corner ф application of formulas of Fresnel (1a-1б) will demand also a precise measurement of the angle of indicence and refraction angle definition already from iterative calculation of the transcendental equation. Though these complexities do not concern to insuperable, number of systematic errors of measuring thus increases.

However it is possible to use the measuring scheme at small angles of incidence in approximation of the angular smallness. We will estimate an error of application of the formula (3) at a diversion from requirements of the normal incidence angle. We will substitute the first 2 terms of decomposition $\cos\varphi$: $\cos\varphi \cong 1 - \varphi^2/2$ in the formula (1), which under condition $\varphi \approx \psi$ transforms to following:

$$\sqrt{R}\,(\varphi) = (n-1) / (n+1 - \varphi^2) \qquad (4)$$

In this case the relative error of measuring «n» under the formula (3) will depend on an angle of indicence as follows:

$$\sqrt{R}/\sqrt{R}\,(\varphi) = (1 - (\varphi^2 / (n+1)) \qquad (5)$$

From here we find the IR as following:

$$n = [1+\sqrt{R}\,(\varphi)\,(1-\varphi^2)] / [1-\sqrt{R}\,(\varphi)] \qquad (5a)$$

For example, for $\varphi = \psi = 1^0$ and $2^0$ relation (5) makes 0.99940 and 0,99885 even at п=1.

The optical plan of measurings in this approach is presented on fig. 1. As the important point of procedure of measurings the select of a mode of the photodetectors providing the linear response on a current within several orders acts. The plane silicon photodiodes FD-24 in a mode of backward bias and at a proper relation between the power supply and the usefull signal (not less than 10) provide linearity of a photoresponse in boundaries of 3 orders. Digitazion of signals was carried out by the digitizedl voltmeters UN70C having an exit in standard RS-232. Capacity of the voltmeter provided recording 4 decimal digits at an error of measuring of 0,1%. At that time registered power of typical He-Ne laser and, accordingly, all detected signals fluctuate in 3-4 sign in the several seconds (without stabilisation), but fluctuations are eliminated on the long-term intervals of averaging in the third sign, the following procedure of calibration of system has been used.

Instead of the sample initially the mirror approximately 100 % reflectivity $R_{cal}$ was set. Then a series of measurings of the power registered by photodetectors FD-A and FD-B. Operation of averaging and evaluation of deviation (variance) was yielded allows to judge reproducibility of results. As the photodetector FD-B registers in this case power impinging on sample of study, the relation of average indications <P (FD-B)> $R_{cal}$ / <P (FD-A)> = M – ( beam splitter transmission) allows on the stage of measurings when instead of mirror $R_{cal}$ set the sample, to spot value of impinging power under indications FD-A and value M. Next values on similar procedure of measuring and processing spotted a reflectivity from an sample of study:

$$R = \langle P\ (FD\text{-}B)\rangle / \langle P\ (FD\text{-}A)\rangle\ M \qquad (6)$$

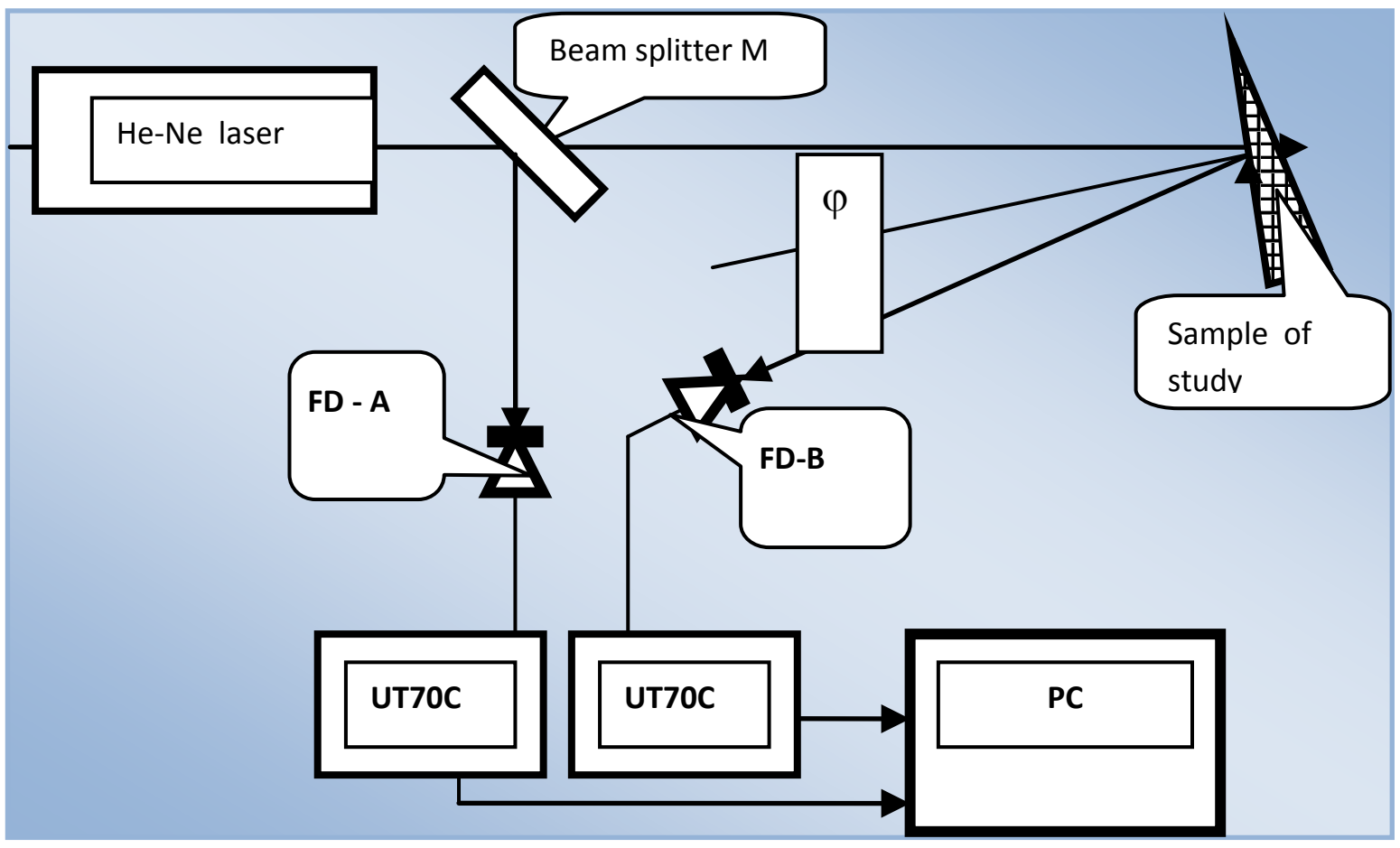

**Fig.1.**

At measuring «n» the transparent films it is necessary to eliminate a reflected beam from the second surface. For the transparent films on a substrate power of a secondary beam is supressed at affinity their IR. In the case of absorbing films the secondary beam cab be supressed also. Level of the dark noise of the given registering plan was at level ≈ 0,2мВ.

Let's consider results of IR measured with scheme fig.1. for glass К8. The sample in the form the optical quality plate and thickness some mm in order to separate in space beams from exterior and interior surfaces. The measured reflectivity of the gauge mirror on a wave length 632,8нм He-Ne laser was equal $R_{cal}$ =0,984. The datas are presented in table 1. The angle of indicence on a sample (and on the gauge mirror) was invariable and equal ≈1,5 $^{0}$. Maintenance of beam geometry for the channel with a photodetector FD-A is important also, but in the case all devices at measurement need not adjustment. Tab. 1. are cited datas for 2 series of measurings.

**Tab. 1.**

| **Glass K8, T ≈$22^0$, $\lambda$ = 632,8нм** | | | | |
|---|---|---|---|---|
| **№№ and calculation** | M | | **Reflectivity** | |
| **n** | **P (FD-A) mV** | **P (FD-B) mV** | **P (FD-A) mV** | **P (FD-**B) mV |
| **1** | 1048,8 | 1171 | 1049,4 | **49,45** |
| **2** | 1050,0 | 1175 | 1048,3 | **49,40** |
| **3** | 1054,9 | 1180 | 1048,6 | **49,39** |
| **4** | 1056,8 | 1181 | 1049,4 | **49,36** |
| **5** | 1061,1 | 1184 | 1050,3 | **49,45** |
| **<….>** | **1055,7** | **1180** | **1049,0** | 49,40 |
| $\sigma_x$ | 0,0039 | 0,0032 | 0,00076 | **0,032** |
| **M** | **1,118** | | **R=0,047** | |
| | | | **R*Rcal/M=0,0438** | |
| | | | **√R*Rcal/M=0,2092** | |
| | | | **n = 1,528** | |
| **Second measuring, n=5** | | | **n=1,518** | |

The analysis of a variance of the gained results in this method shows that for reproducibility of of measuring IR in 4th sign the number of measurings shoulb be increased.

### 3. MEASURING IR WITH BREWSTER ANGLE

Owing to traversal structure of electromagnetic fields a reflectivity of light polarized in a plane of incidence (1б) at $(\varphi_B + \psi) = \pi/2$ aspires to zero, what is followed $\varphi_B$+arcsin $(\sin\phi_B/n) = \pi/2$, which solution is value of Brewster angle as function of IR:

$$\phi_B=\arcsin (n^2/n^2+1)^{0,5}) = \mathrm{arctg}\,(n) \qquad (7)$$

The formula (7) specifies a possible method of measurement IR of different materials. Measuring procedure «n» to be reduced to Brewster angle measuring, need not precise measurings of testing beam power and, consequently, can provide measuring with necessary accuracy.

Here pertinently to note that presence of the reflected power with elliptic polarization at a Brewster angle specifies in diversions from Fresnel theory. This contradiction removes at introduction in the theory of Relay hypothesis about the intermediate stratum in the thickness a lot of a smaller wave length /5/. The calculated values of a thickness of the intermediate stratum within the limits of the modified Fresnel theory resulted in value of stratum thickness for different fluids $\approx 10^{-3}\lambda$. The modified theory does not render influence on value of mass coefficient of IR of the basic material with use of primary Fresnel theory.

Let's analyse requirements of correct application of a relation (7) with the given purpose. According to (7) measuring of a Brewster angle on a minimum of reflection of a testing laser beam with a margin error 1 angular minutes are provided with measurement “n” with a margin error in 4th digit. Thus it is supposed that measuring is spent with a beam of 100 % a degree of the linear polarization: the depolarisation leads to bias of a required minimum of reflection with entering of a corresponding error for Brewster angle measuring. Similarly to depolarization the minimum of reflection is shifted by a lateral angle α - misfit of planes of incidence and polarization. The full reflectivity of randomly polarized radiation as lateral angle function can be written down following expression [5]:

$$R(\alpha) = R(\parallel)\cos^2(\alpha) + R(\perp)\sin^2(\alpha) \qquad (8)$$

For randomly polarized light medial value in all polarization directions $\langle\cos^2(\alpha)\rangle = \langle\sin^2(\alpha)\rangle = 1/2$; for light with 100 % linear polarisation at $\alpha = 0^0$ power of reflected light is determined by the first term (8) having the reflection at a Brewster angle; at $\alpha = 90^0$ - by the second term in (8 ) which does not have a extremum on the similar angular dependence. For any lateral angle $\alpha \neq 0 \neq 90^0$ shape of the resultant curve of reflection is spotted by the total of terms in (8), thus the minimum will be biased towards smaller angles with increase α. The case with light of a restricted degree of polarisation also can be reduced to expression of a type (8). Designating a degree of polarisation «p» expression (8) is rewritten as:

$$R(p, \alpha) = R(\parallel)(p + \cos^2(\alpha))/2 + R(\perp)(1 - p + \sin^2(\alpha))/2 \qquad (9)$$

At p=1, α =0 the case is realised ideal for measuring of the true minimum. At p=0,5 and α =0 we have a case similar to application of randomly polarized radiation. Below we will estimate admissible for measuring with the given error the polarisation degree. Fig. 2 represents pictorial dependence of reflection for two extreme values of a lateral angle and p=1 and a reflection derivative on an angle of indicence ϕ for α =0 and n=1,505. Equality the derivative to zero on an indicence angle allows to find with the given accuracy a standing of a minimum and its sensitivity to change of IR magnitude: so changes IR in 4th digit: 1,501; 1,502; 1,503; are accompanied by following values of Brewster angles: 0,9831; 0,9834; 0,9837 radian.

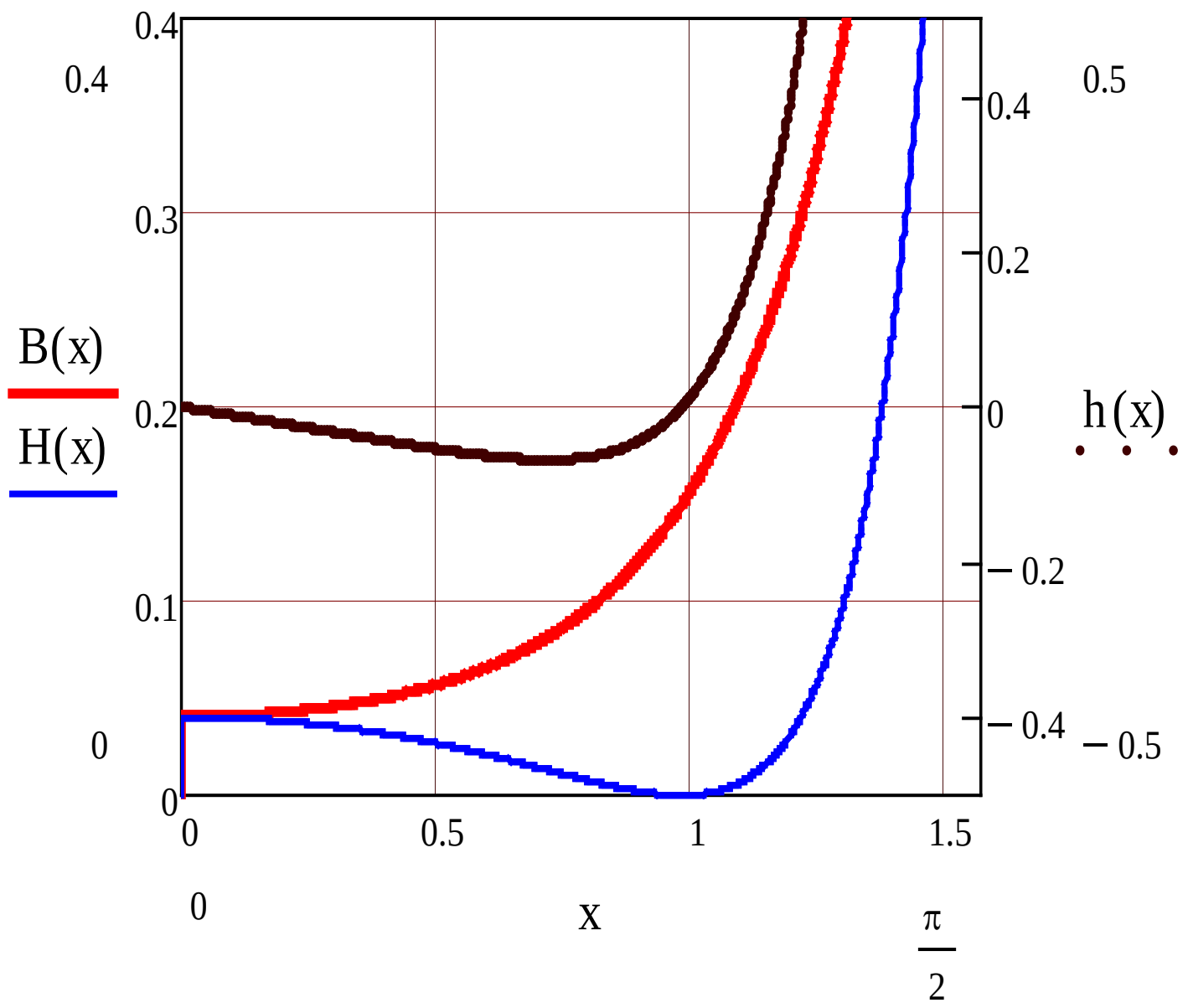

Fig.2. Angular dependence of reflexion of light with p=1 for α =0 (blue), α =90$^{0}$ (red) and derivative reflections on an angle of indicence x=ϕ (black, a radian).

For definition of admissible deviations from the requirement to have p=100 % and α =0 we will calculate corresponding angles of displacement of minimum H (x) on fig. 2.равном 0,9837 for value “n” =1,503. Effects of calculation in a pictorial view are presented on fig. 3. They show that only at a degree of polarisation of 99 % and above minimum shift practically disappears.

Though the true Brewster angle on a reflection minimum is oserved only at 100% of polarization degree and at a zero lateral angle, measuring of an angular standing of a minimum of reflected light with a margin error of 1 angular minute is reached at practically feasible requirements for these operations. It is qualitatively visible that in this measuring it is necessary to apply a beam of light with a degree of polarization not below 99 %. An error of measuring of 1 angular minits causes an error of measuring of IR in 4th digit.

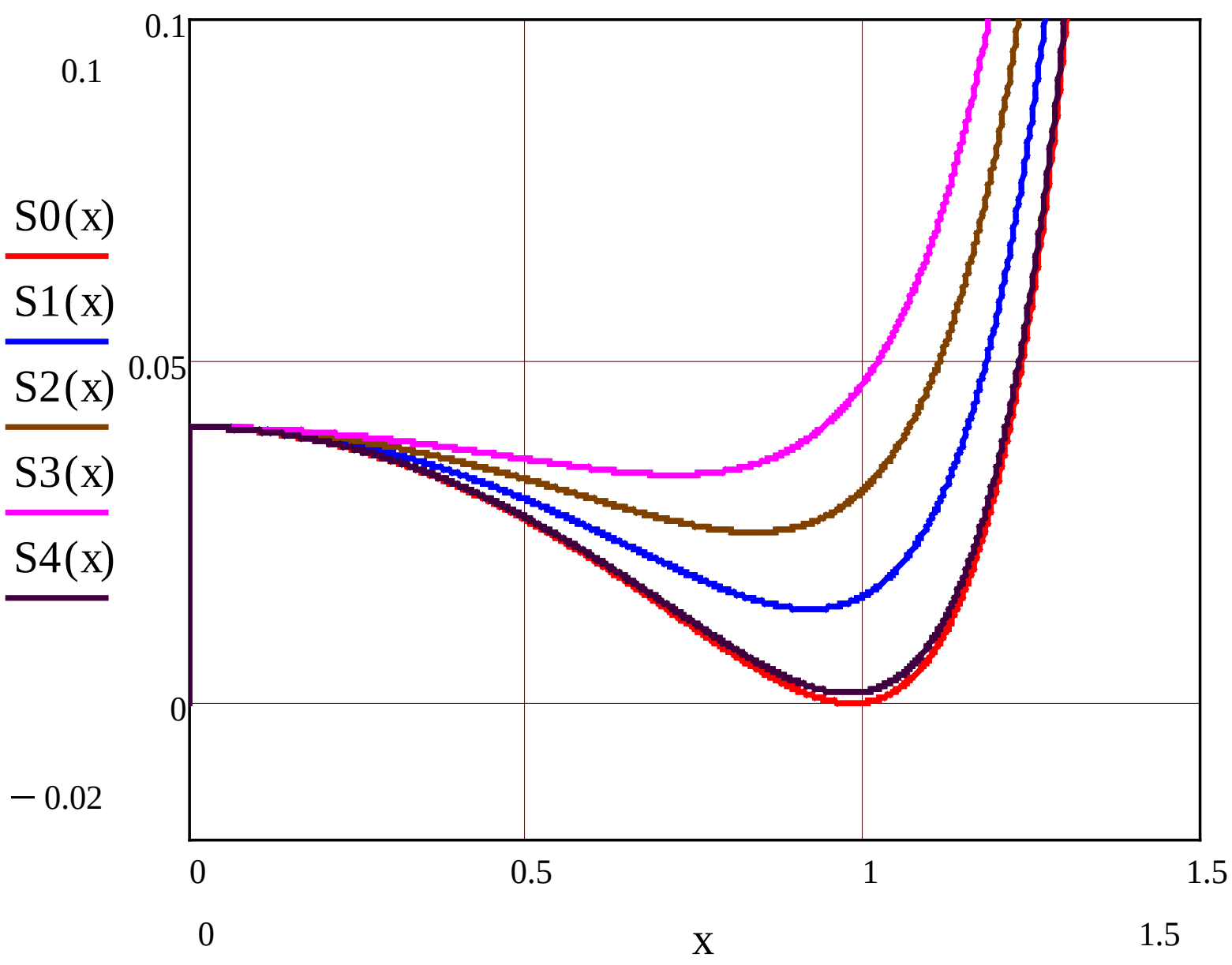

Fig. 3. A standing of a false Brewster angle in a case of light with polarization degree of smaller 100 %:
S0=100 %-red
S4=99 % black
S3=70 % pink
S2=80 % brown
S1=90 % dark blue

Application in this measuring of lasers with the continuous radiation and with the linear polarisation of radiation easily satisfies to the requirement on a degree of polarization more than 99 %.

Misfit of a polarization plane and plane of incidence can be viewed to similarly reduction of a degree of polarization of a measuring beam of light. In that case misfit quantity $0{,}5^0$ makes the same effect as decrease in a degree of polarisation from 100 % to 98,2 % that also feasable in practice.

Divergence of a measuring beam also should meet certain requirements of accuracy of an angular position of a Brewster extremum. But as the angular width of a Brewster reflection minimum of makes some degrees and essentially exceeds angular divergence of typical lasers (several angular minits) the error imported by divergence of a measuring beam is neglected small. Moreover, search of an angular position of a minimum is accompanied by measuring not one point, and their whole plurality within several degrees that is accompanied by averaging and removal of this error at all.

The following prominent feature of Brewster angle measuring is interfaced to difficulties of limitedly low power recording around of a minimum: at low power of a measuring beam the noise level in this limit can exceed the useful signal. This complication is overcome by increase of measuring beam power or by application of procedure of restoration of a missing part of dependence in a minimum on the measured points by regression analisys with polynomes of a degree necessary for reaching of a correlation factor ≈1. The termed procedure is accessible in known applied mathematical programs.

The monitoring part and light sources in this procedure of IR measuring remained former as well as presented on fig. 1. However a photodetector FD-B and the sample are erected on goniometer GUR--4, having accuracy of angular rotation 1 minute. Besides the goniometer gyration system provided possibility of 2-fold angular rotational displacement of a photodetector at unitary rotational displacement of the sample with maintenance of an immovability of a laser beam on a photodetector plate. Installation of zero of a frame of angular reference was yielded by an autocollimation of a laser beam at its normal incidence on an sample surface with a margin error to 1 angular minut.

Now with known parameters of a used laser beam is possible to start IR measurings of prepared materials. As the procedure does not discover the rigid requirements to stability of power of generation, tracing of named parameter during measuring is not binding. On fig. 4. results of measuring and processing on an example of heavy flint sample are presented. It was seen from the data for fig. 4. that in a neighbourhood of a Brewster angle minimum is not spotted due to excess of noise over the signal. However after polynomial fit operation ( 3 degrees) the measuring curve spreadsheet contains the required singular minimum (it is spotted at step-type behavior of digitation in shares of minutes). The analytical shape of the found polynomial is presented in the same place. The Brewster angle determined on a minimum is equal $59{,}12^0$ corresponds IR = 1,672 with probability 99 % on a confidential interval 0,71-0.701.

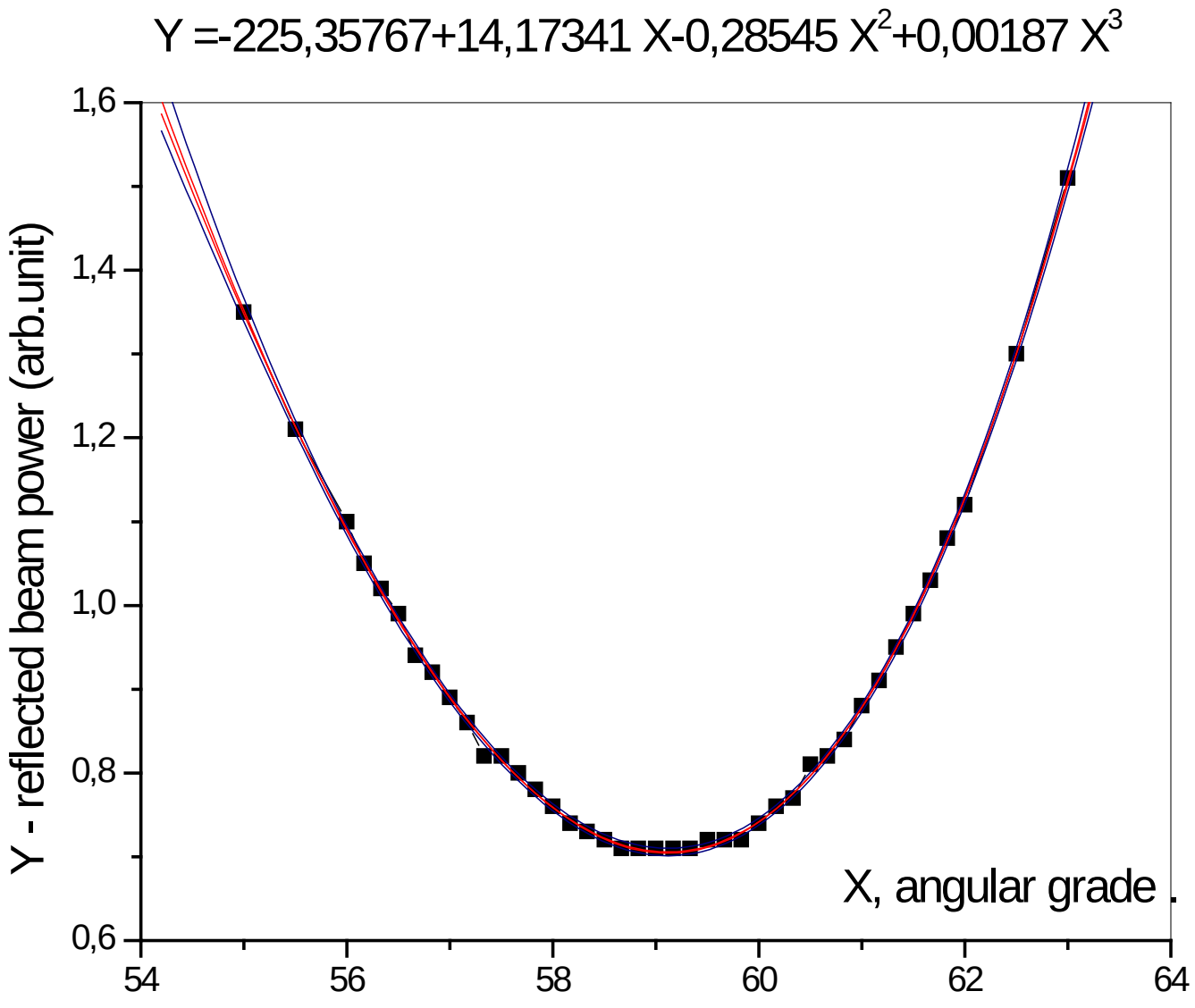

Fig. 4. Dependence of reflected light in a neighbourhood of a corner of Brjusteradlja of heavy flint on a wave length 632,8нм, a degree of polarisation p=99,8 %, $\alpha < 1^0$

For an illustration of possibility of application of a method to light absorbing (on a measuring wavelength) materials we will spend IR measuring and processing of "grey" glass filter NS-7.

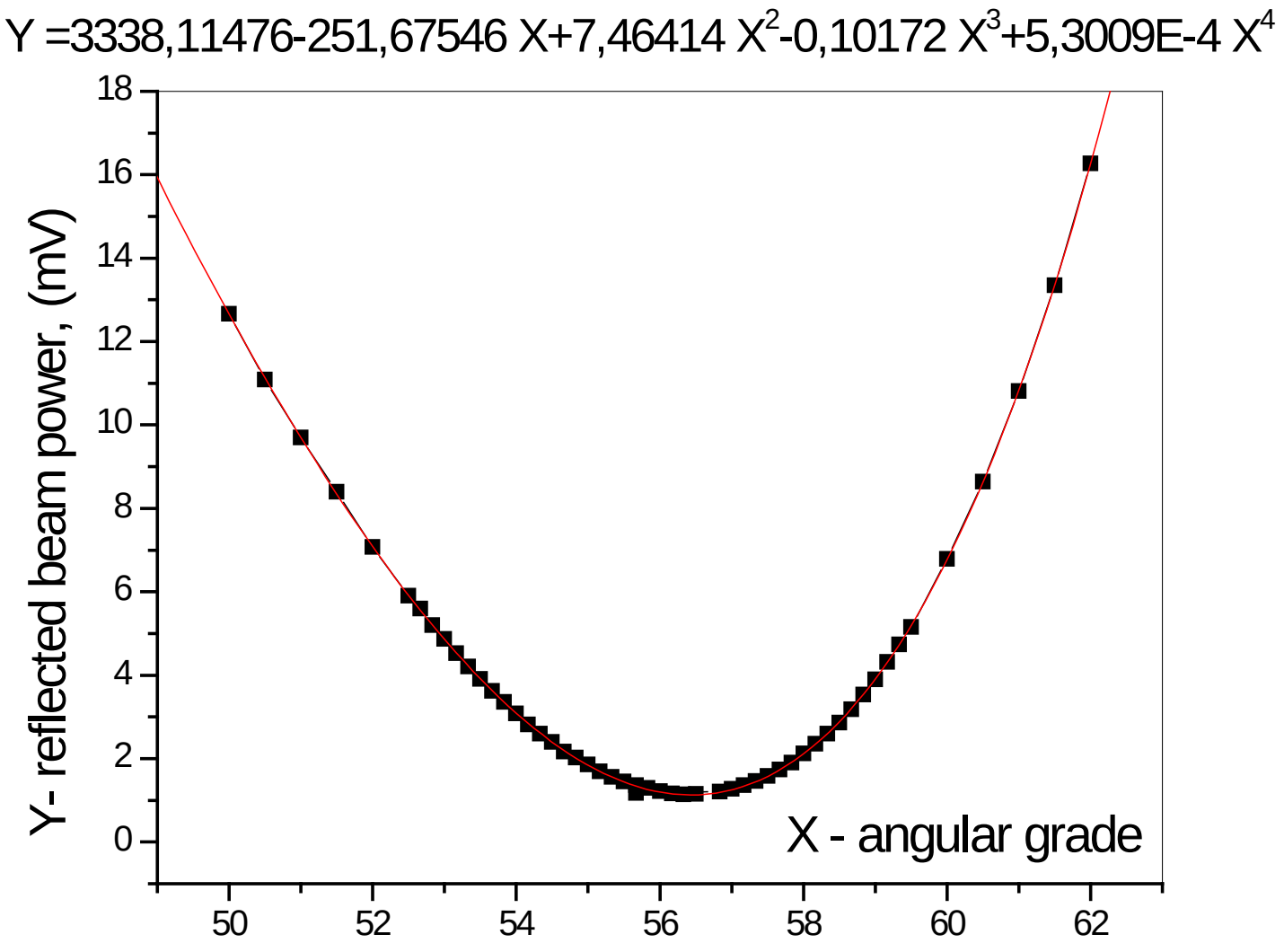

Fig. 5. Dependence of reflected light in a neighbourhood of a Brewster angle for “grey” light filter NS-7 and determination of its real part

Fitting of the detecting points that corresponds region signal/noise>1 is yielded by a polinomial of 4th degree for reaching of a correlation factor 0,99. Spotted on a Brewster angle (56,42±0,016) $^0$ IR it has appeared equal n = 1,506. The systematic error 1 angular minute is related to above scored method of the zero assignment of an initial standing of sample on a rotary platform.

At use of the given procedure for IR determination of absorbing samples it is necessary to make some remarks. Really, IR in similar cases has complex value. The relation of magnitudes of the real and imaginary parts IR spots presence or disappearance Brewster angle and, accordingly, possibility of the given procedure for its determination. Till now we dealt with the transparent materials with the real IR. Metals and films of organic dyes concern strongly absorbing materials in case of coincidence of frequencies of radiation to allowed absorption bands. It is known that R. Wood managed to find a Brewster angle for argentum in its UF transparency region /5/. We will estimate possibilities of the considered procedure with reference to the coloured by organic dye solutions and polymeric compounds.

Complex wave vector of passage of a plane wave in the immersing medium:

$$k(\lambda,n) = k_r + jk = (2\pi/\lambda)(n + j\varkappa) = 2\pi n/\lambda + j\,2\pi\varkappa/\lambda = 2\pi n/\lambda + j\alpha \qquad (10)$$

ϰ and α – an exponent and absorption coefficient, λ and n- a wavelength and real IR . The relation between absorption coefficient and an absorption exponent $\lambda\,\alpha/2\,\pi = \varkappa$ allows to estimate the quantitative boundary between feebly and strongly absorbing materials (tab. 2.)

Below the boundary scored by yellow colour where the field of appreciable effect of absorption coefficient on IR and reflection of a material begins and, accordingly, there is a decrease in angular and polarization dependence of reflectivity of incident radiation from material surface.

Tab. 2. (Initial parameters of calculation: КП1,555, λ=500нм)

| №№ | Absorption coefficient α ($cm^{-1}$) | Absorption exponent ϰ | The real part IR: $\sqrt{n^2 - \varkappa^2}$ | Imaginary part IR $\sqrt{2n\varkappa}$ | Reflectxion at normal incident, % |
|---|---|---|---|---|---|
| 1 | 1 | $7{,}962*10^{-6}$ | 1,555 | $4{,}967*10^{-3}$ | 4,7 |
| 2 | 10 | $7{,}962*10^{-5}$ | 1,555 | $15{,}73*10^{-3}$ | 4,7 |
| 3 | 100 | $7{,}962*10^{-4}$ | 1,555 | $4{,}967*10^{-2}$ | 4.7 |
| 4 | 1000 | $7{,}962*10^{-3}$ | 1,555 | $15{,}73*10^{-2}$ | 4,7 |
| 5 | 10000 | $7{,}962*10^{-2}$ | ≈1,555 | $4{,}967*10^{-1}$ | 5,4 |
| 6 | 100000 | $7{,}962*10^{-1}$ | 1,150 | 1,573 | 13,22 |
| 7 | 495222 | 1,555 | 0 | 2,199 | 31,72 |

Correspondently scored in tab. 2. of absorption (α) above yellow boundary specifies the field of direct applicability of Brewster method of measuring IR for feebly absorbing materials. Region below the shown boundary concerns to material where even observed minimum reflection and Brewster angle is connected with complex IR. Determination the complex IR on the Brewster angle measurement will be consider in our next work.

It is well known that for many applications IR of thin films (quantity and dispersion) /6,7/ has powerful value. It is known also that for film thickness equal and above to the measuring wavelength of refractometer IR does not lose the physical sense. In a context of operations spent by authors on photopolymeric holographic composites /7/ the particular interest represents their IR. Holographic diffraction elements on the basis of photopolymeric films, as a rule, contain substrates. If IR polymetric film and substrate are close, the affinity removes necessity of elimination of the second beam contribution, intolerable at procedure measurement on a Brewster angle. On fig. 6. results of a such measuring IR of a photopolymeric film and the analytical shape of a curve of reflection in the form of polynomial of 4th order is presented. To find a point of a Brewster minimum with possible a split-hair accuracy 1000 points in a gamut of the measuring have been chosen at different angles. The quantity of the observational points for restoration of dependence by regression fitting procedure is required 10 times less. Calculated data for the case are presented on fig.6. below.

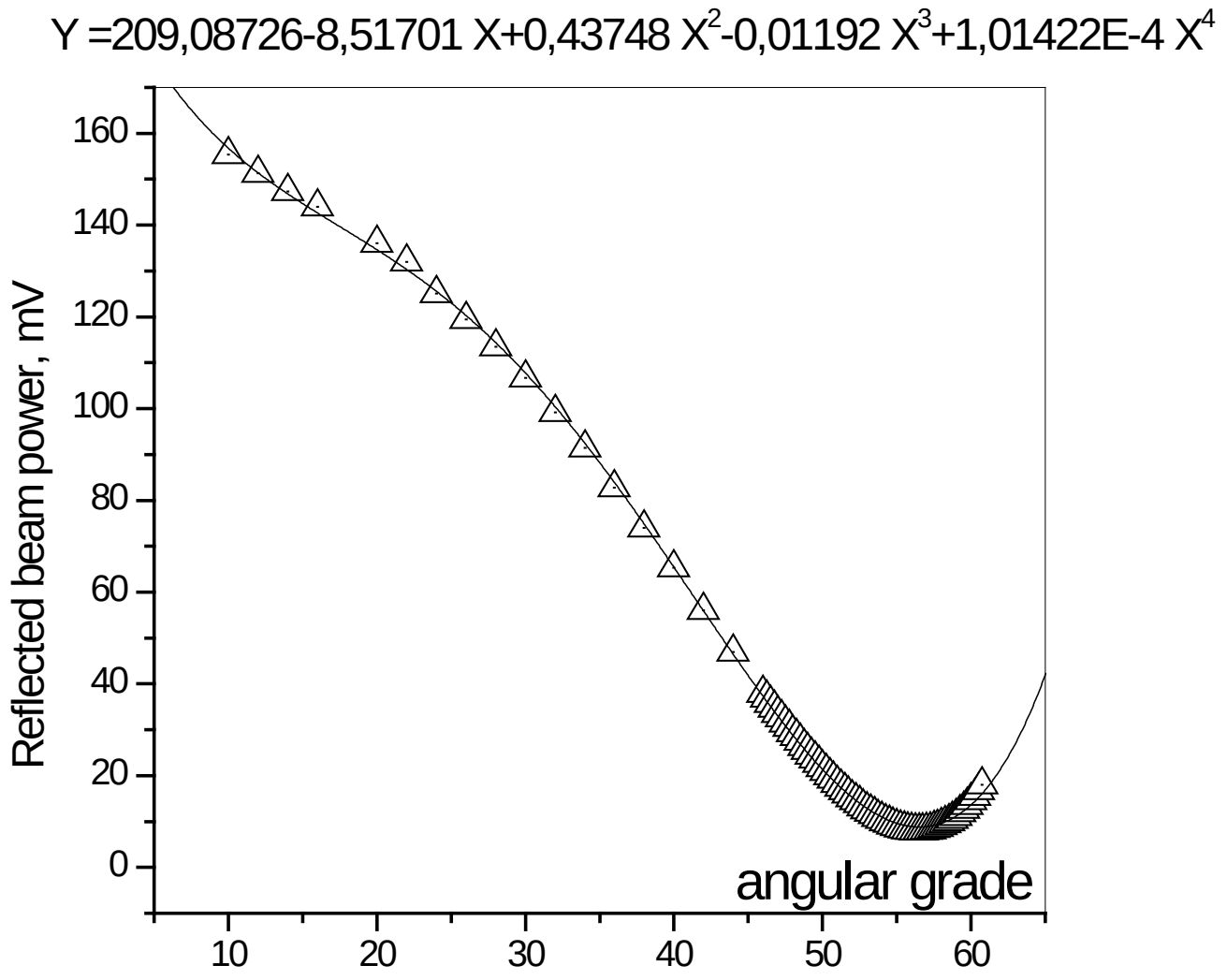

Fig.6. A reflection minimum is in a point of 56,559 angular degrees and corresponds IR n=1,514 on λ = 632,8 nm.

4. A comparative estimate of methods and conclusions

It is clear that both featured methods measuring of real IR principally can provide identical results and errors. Both methods of the refractometry in reflected light are characterised by the aprioristic advantage concerning traditional thanks to possibility to measure IR for various condensed mediums only with one prepared surface of optical quality.

Errors of the first method are stipulated with measuring of beam powers; errors of the second method are stipulated with the complications arising at a finding of the true Brewster angle. Errors of laser beam power measuring are caused by temporary instability of radiation of the serial devices, spectral sensitivity of semiconductor detector and thermal noises. Though these restrictions does not set the unmovable limit to achieve stability in 3-4 sign to decrease this error by statistical processing it is difficultly. However, on physical requirements, the method of measuring IR in reflected light at almost normal incidence is accompanied the smaller number of requirements to spend correct measuring in compare to Brewster angle method. Ellipsometry is well-known example of development of refractometry in reflected light for study of the various strongly absorbing materials.

High-accuracy refractometry based on Brewster angle measuring needs testing beam with degree of polarization (> 0,98), exact installation of a small lateral angle (<$1^0$) and a zero angle of readout of the reflection plane concerning which Brewster angle measuring is yielded. Though stability of beam power for this method is less significant, stable power also is important at finding reflection minimum (Brewster angle) directly or in a case of observational point fitting at signal/noise relation <1.

In spite of that refractometry based on Brewster angle measurement on our opinion nowadays has good conditions for development.